\documentclass[journal,comsoc]{IEEEtran}

\usepackage[T1]{fontenc}

\ifCLASSINFOpdf
\else
\fi

\usepackage{amsmath}

\usepackage[cmintegrals]{newtxmath}

\usepackage{graphicx}
\usepackage{subfigure}
\usepackage{multirow}
\usepackage{color}
\usepackage{enumerate}
\usepackage{framed}

\begin{document}

\title{The Decoding Success Probability of Sparse Random Linear Network Coding for Multicast}

\author{WenLin~Chen, Fang~Lu ~\IEEEmembership{Member,~IEEE,} and Yan~Dong ~\IEEEmembership{Member,~IEEE}

\thanks{The work was supported by the National Nature Science Foundation of China (91538203).
}
\thanks{W. Chen, F. Lu and Y. Dong are with the School of Electronic Information and Communications,
Huazhong University of Science and Technology, Wuhan 430074, China
(e-mail: wenlinchen@hust.edu.cn,lufang@hust.edu.cn; dongyan@hust.edu.cn).}}

\markboth{IEEE TVT}%
{Submitted paper}

\maketitle

\begin{abstract}
Reliable and low latency multicast communication
is important for future vehicular communication. Sparse random
linear network coding approach used to ensure the reliability
of multicast communication has been widely investigated. A
fundamental problem of such communication is to characterize
the decoding success probability, which is given by the probability of a sparse random matrix over a finite field being full rank. However, the exact expression for the probability of a sparse random matrix being full rank is still unknown, and existing approximations are recursive or not consistently tight. In this paper,
we provide a tight and closed-form approximation to the probability of a sparse random matrix being full rank, by presenting the explicit structure of the reduced row echelon form of a full rank matrix and using the product theorem.
Simulation results show that our proposed approximation is of high accuracy regardless of the generation size, the
number of coded packets, the field size and the sparsity, and tighter than the state-of-the-art approximations for a large range of parameters.
\end{abstract}

\begin{IEEEkeywords}
Sparse random linear network coding, sparse matrices, decoding success probability, multicast communications.
\end{IEEEkeywords}

\IEEEpeerreviewmaketitle

\section{Introduction}\label{section1}
In future vehicular communication, multicast communication with high reliability and low latency is critical. For example, in automated overtake use case which requires cooperation among vehicles to avoid a collision, safety-of-life data needs to reach all participants with ultra-high reliability and ultra-low latency \cite{5G-PPP}. Traditionally, the reliability of multicast communications is ensured by a digital fountain approach \cite{2002Byers}, which is typically realized by Luby transform or Raptor codes.
However, these codes achieve their capacity only if the number of source packets per generation is large, which could be an issue for delay-critical data. As an alternative to fountain codes, Random Linear Network Coding (RLNC) approach, which is used to ensure the reliability of multicast, has attracted extensive research attentions \cite{2017Chatzigeorgiou,2018Tsimbalo}. Unfortunately, a key concern of RLNC approach is its relatively high decoding complexity. Hence, this paper adopts Sparse Random Linear Network Coding (SRLNC) approach to ensure reliability of multicast, as in \cite{2016Tassi,2018Brown}.

In SRLNC for multicast, the sender splits data into the generations, each of which consists of $n$ source packets. During each generation, the sender injects a stream of coded packets into the network. Each coded packet is obtained by linearly combining the $n$ source packets over a finite field $F_q$ of order $q$, where the coefficients choose the zero element with probability $p_0$ and each non-zero element with equal probability $(1-p_0)/(q-1)$. A receiver recovers $n$ source packets as soon as it collects $n$ linearly independent coded packets. In such multicast network, a key performance metric is the probability of a receiver recovering $n$ source packets, which is referred as the decoding success probability\footnote{In SRLNC, the decoding matrix at a receiver is a sparse random matrix over $F_q$. Hence, the terms ``SRLNC'' and ``sparse random matrix'' are indiscriminately used in this paper.}.
To the best of our knowledge, an exact expression for the decoding success probability of SRLNC, as a function of the generation size, the number of coded packets, the field size and the sparsity, is still unknown.

Several works have studied the nonsingular probability or the rank distribution of a sparse random matrix over $F_q$ under the asymptotic setting.
Charlap \textit{et al.} \cite{1990Charlap} proved that, as long as the distribution of the entries of the matrix is not concentrated on any proper affine subspace of $F_q$, the asymptotic nonsingular probability of a random $n\times n$ matrix over $F_q$ is the same as for uniform entries. A more generic conclusion was presented in \cite{2001Kahn}.
Cooper \cite{2000-10-Cooper} proved that, conditioned on the event that the matrix has no zero rows or columns, the asymptotic rank distribution of a sparse random $(n-s)\times n$ matrix over $F_q$ is the same as for uniform entries, where s is a nonnegative integer.
By introducing the concept of zero pattern of the random matrix, Li \textit{et al.} \cite{2011-06-Li,2011-07-Li} derived an upper and lower bound on the singularity probability and the rank distribution of a sparse random square matrix over $F_q$.
However, these results are proved only for infinite matrix size ($n\to \infty$) \cite{1990Charlap,2001Kahn,2000-10-Cooper} or infinite finite field size ($q\to \infty$) \cite{2011-06-Li,2011-07-Li}. In practical SRLNC scenarios, the matrix size and the finite field size can not be very large. Therefore, it is desirable to obtain the expression applied to any matrix size and any finite field size.

Many works have devoted to characterize the performance of SRLNC under the non-asymptotic setting. Based on Absorbing Markov Chain (AMC), Garrido \textit{et al.} \cite{2017Garrido} characterized the performance of SRLNC in terms of the decoding success probability and the average number of transmissions required to decode a generation. Unfortunately, the transition probability of AMC relies on Monte Carlo simulation results and AMC model therein is not analytical.
Based on the Stein-Chen method, Brown \textit{et al.} \cite{2018Brown} derived an approximation to the probability of a sparse random $n\times m$ matrix over $F_q$ being full rank. However, the approximate expression is recursive and is not tight for large $q$ and $p_0$. Based on linear dependency of a matrix, Sehat \textit{et al.} \cite{2019Sehat} derived an approximation to the probability of a sparse random matrix over $F_q$ being full rank. Then based on this probability, they presented a recursive model for the rank distribution of sparse matrices. However, they do not consider the correlation between linear dependencies of a matrix, hence their approximation to the probability of a sparse random matrix over $F_q$ being full rank is very loose, thereby impacting on the tightness of the rank distribution. In addition, this recursive model is essentially a Markov chain, which does not always provide closed-form expression.

In addition, there are several works based on the results of \cite{1997Blomer}.
Bl$\ddot{o}$mer \textit{et al.} in \cite[Th. 6.3]{1997Blomer} derived an upper-bound on the probability of a $n$-dimensional vector being dependent of other $i$ linearly independent vectors, which is denoted as $p(i,n)$, and hence obtained a lower bound on the nonsingular probability of a sparse random $n\times n$ matrix over $F_q$.
Feizi \textit{et al.} in \cite[Lem. 1]{2014Feizi} also obtained an upper-bound on the probability $p(i,n)$, which can be proven to be equivalent to \cite[Th. 6.3]{1997Blomer} for $p_0 \ge 1/q$.
By combining \cite[Th. 6.3]{1997Blomer} and AMC, Tassi \textit{et al.} \cite{2016Tassi} characterized the performance of SRLNC in terms of the average number of coded packet transmissions needed to recover the source message. By extending \cite[Th. 6.3]{1997Blomer}, Khan \textit{et al.} \cite{2016Khan} derived an upper and lower bound on the singular probability of sparse random matrices over $F_q$.
However, the upper-bound on the probability $p(i,n)$ presented in \cite[Th. 6.3]{1997Blomer} is not tight for $n > 10$ and $p_0\ge 0.7$, which has an impact on the tightness of \cite{2014Feizi,2016Tassi,2016Khan}.
Recently, based on the Reduced Row Echelon Form (RREF) of a full rank matrix, we derived in \cite{2019Chen} an approximation to the probability $p(i,n)$, which is much tighter than \cite[Th. 6.3]{1997Blomer}. Then we established an exact expression for the rank distribution of sparse random matrices over $F_q$ as a function of $p(i,n)$.
However, the correlation between the entries of a $n$-dimensional vector contained in a $i$-dimensional subspace is not considered. As a result, the approximation to $p(i,n)$ is not tight for small $n$ and large $p_0$. Furthermore, the established exact expression for the rank distribution as a function of $p(i,n)$ is of high computational complexity and approximated, which further reduces the tightness of the rank distribution.

In this paper, we focus on the decoding success probability and the main contributions are summarized as follows:

\begin{itemize}
\item We improve the analysis for the probability $p(i,n)$ in \cite{2019Chen}, and derive a generally tighter and closed-form approximation to the probability $p(i,n)$. First, we present the explicit structure of the RREF of a full rank matrix, which allows the dependent entries of the RREF of a full rank matrix to be decomposed into independent or approximately independent components. Second, we prove that the non-zero coefficients of a linear combination of sparse random variables have no effect on the distribution of that linear combination. This insight allows the first $i$ entries and the last $n-i$ entries of a $n$-dimensional vector contained in a $i$-dimensional subspace to be decoupled.
\item We derive a low complexity and exact expression for the probability of a sparse random matrix over $F_q$ being full rank as a function of $p(i,n)$, which needs no any approximations.
\item Unlike existing approximations \cite{2018Brown,2019Chen,2019Sehat} which are recursive or not consistently tight, we propose a novel approximation to the probability of a sparse random matrix over $F_q$ being full rank, which is closed-form and of high accuracy regardless of the generation size, the number of coded packets, the field size and the sparsity. Our proposed approximation is tighter than \cite{2018Brown,2019Chen,2019Sehat} for a large range of parameters.
\end{itemize}

The rest of the paper is organized as follows. In Section \ref{section2},
we describe the considered system model. In Section III,
a novel approximation to the probability of a sparse random matrix over $F_q$ being full rank is derived.
In Section IV, we validate the accuracy of our proposed approximation by Monte Carlo simulations and compare it with the state-of-the-art approximations proposed in previous works. Finally, we draw our conclusions and future work in Section V.

\section{System model}\label{section2}
We consider
a multicast network where a sender transmits data to multiple receivers. We assume that each link from the sender to a receiver is lossy and characterized by a packet erasure rate $\epsilon$. In order to ensure the reliability of multicast communications, the sender transmits data encoded by SRLNC approach.

The sender splits data into the generations, each of which consists of $n$ source packets $\{x_1,x_2,\cdots,x_n\}$. Each source packet $x_k,~k=1,2,\cdots,n$ consists of $L$ elements from $F_q$. During each generation, the sender injects a stream of coded packets $\{y_1,y_2,\cdots,y_N\}$ into the network.
A coded packet $y_j,~j=1,2,\cdots,N$ is defined as
\begin{equation*}
y_j=\sum_{k=1}^{n}g_{k,j}x_k,
\end{equation*}
where $g_{k,j}\in F_q$ is referred as the coding coefficient, and the vector $(g_{1,j},g_{2,j}\cdots,g_{n,j})^{T}$ is referred as the coding vector.
Using the matrix notation, the encoding process can be expressed as
\begin{equation*}
(y_1,y_2,\cdots,y_N)=(x_1,x_2,\cdots,x_n)G,
\end{equation*}
where $G=(g_{k,j})$ is an $n\times N$ random matrix over $F_q$.
The coding coefficients $g_{k,j}$ are independently and randomly chosen from $F_q$ according to the following distribution:
\begin{equation}\label{equation1}
Pr\{g_{k,j}=t\}=
\begin{cases}
p_0, &t=0 \\
\displaystyle\frac{1-p_0}{q-1}, &t\in F_q\setminus \{0\}
\end{cases}
\end{equation}
where $0\le p_0 \le 1$ is referred as the sparsity of the code. The RLNC scheme refers to $p_0=1/q$ (i.e., the coding coefficients are uniformly chosen from $F_q$), while SRLNC scheme is characterized by $p_0>1/q$.

It is worth mentioning that there is the possibility of the sender generating zero coding vectors since the coding vector is randomly generated. From a perspective of real implementation, zero coding vector should be not transmitted since it results in the transmission overhead. However, this paper includes the transmission of zero coding vectors as in \cite{2016Tassi,2018Brown}, thus the performance modeling is tractable and more general.

Due to packet erasure, each receiver will receive a subset of transmitted coded packets.
Let $m$ denote the number of coded packets received by a receiver.
The receiver constructs an $n\times m$ decoding matrix $M$ with $m$ received coded packets. Obviously, the matrix $M$ is obtained from $G$ by deleting the columns corresponding to erased coded packets.
The receiver can recover $n$ source packets if and only if the rank of $M$ is equal to $n$. We now define the decoding success probability $P(\epsilon)$ as the probability of a receiver successfully recovering $n$ source packets. Note that although packet erasures affect the number of coded packets required by a receiver to recover a generation, it is independent of the random choice of coding vectors \cite{2011Trullols-Cruces}.
Therefore, $P(\epsilon)$ can be expressed as follows:
\begin{equation*}
P(\epsilon)=\sum_{m=n}^{N}\binom{N}{m}(1-\epsilon)^{m}\epsilon^{N-m}P(n,m),
\end{equation*}
where $P(n,m)$ is the probability of a $n\times m$ decoding matrix $M$ being full rank. In the remainder of this paper, we focus on the probability $P(n,m)$.

\section{Performance Analysis}
\label{section3}
\subsection{Previous Results Analysis}
Let $M_{m\times n}$ be the transpose of the matrix $M$. In \cite{2019Chen}, authors found that $M_{m\times n}$ is full rank if and only if there are $n$ rows increasing the rank and $m-n$ rows maintaining the rank in $M_{m\times n}$. In fact, the probability of a row maintaining the rank is $p(i,n)$.
Then the probability of $M_{m\times n}$ being full rank is given by
\begin{equation*}
\begin{split}
&Pr\{rk(M_{m\times n})=n~|~M_{m\times n}~\text{has no zero columns}\}\\
=&\prod_{i=0}^{n-1}\big(1-p(i,n)\big)\\
&\sum_{r_1=0}^{n}p(r_1,n)\sum_{r_2=r_1}^{n}p(r_2,n)\cdots\sum_{r_{m-n}=r_{m-n-1}}^{n}p(r_{m-n},n)
\end{split}
\end{equation*}
In order to calculate above equation, \cite{2019Chen} provided an approximation to $p(i,n)$ based on the RREF of a full row rank matrix, and simplified the embedded sums whose computational complexity are $O(n^{m-n})$ by the properties of Gauss Coefficient.

However, there are three approximations impacting the tightness of \cite{2019Chen}:
\begin{enumerate}[i)]
  \item According to the RREF of a full row rank matrix, it can be proved that $p(i,n)$ is equal to the probability of a row $h_{i+1}$ being that the first $i$ entries are arbitrary and the last $n-i$ entries are uniquely determined by the first $i$ entries. Obviously, the last $n-i$ entries of $h_{i+1}$ are not independent since they have common components. However, \cite{2019Chen} assumed that the last $n-i$ entries of $h_{i+1}$ are independent and approximated the joint distribution of the last $n-i$ entries of $h_{i+1}$ by the product of the marginal distributions of them,
  \item In order to calculate the marginal distributions of the last $n-i$ entries of $h_{i+1}$, it is necessary to obtain the distribution of each entry of the RREF of a full row rank matrix. In \cite{2019Chen}, authors provided an approximation to the distribution of each entry of the RREF of a full row rank matrix, based on Gauss elimination. However, if the explicit structure of the RREF of a full row rank matrix is known, this step is not necessary because the entries of the RREF of a full row rank matrix are not independent and need to be further decomposed into independent components,
  \item In order to simplify the embedded sums by using the properties of Gauss Coefficient, the approximate expression for $p(i,n)$ was further approximated by its lower bound. This further reduces the tightness of \cite{2019Chen}.
\end{enumerate}

In this paper, in contrast to \cite{2019Chen}, we improve the proof at two aspects:
i) Based on the product theorem, we establish a lower complexity relation between the probability $P(n,m)$ and the probability $p(i,n)$. This avoids unnecessary further approximation to $p(i,n)$ caused by using the properties of Gauss Coefficient,
and ii) Based on standard linear algebra, we present the explicit structure of the RREF of a full row rank matrix. This means that the entries of the RREF of a full row rank matrix can be further decomposed into independent components, and therefore enables us to significantly mitigate the impact of the dependence between the last $n-i$ entries of $h_{i+1}$.

\subsection{The Probability of A Row Being Linearly Dependent}
In this subsection, our goal is to derive the probability $p(i,n)$. Before we give a derivation, we need some auxiliary lemmas.
In this paper, the notations of multiplication and addition in the finite field are same as in the real field without ambiguity.

The following lemma reveals that if the random variables are i.i.d. with (\ref{equation1}), the sum of such random variables still follows (\ref{equation1}) but with different $p_0$.

\newtheorem{lemma}{Lemma}
\begin{lemma}\label{lemma1}
Let $X_1,X_2,\cdots,X_k$ be independent random variables over $F_q$ and identically distributed with (\ref{equation1}). Then the distribution of the sum of $X_1,X_2,\cdots,X_k$ is
\begin{equation}
\begin{split}
&Pr\{X_1+X_2+\cdots+X_k=t\}\\
=&\beta_k(p_0,t)\\
=&\begin{cases}
\frac{1}{q}\Big(1+(q-1)\big(1-\frac{q(1-p_0)}{q-1}\big)^k\Big),&t=0 \\
\frac{1}{q}\Big(1+(-1)\big(1-\frac{q(1-p_0)}{q-1}\big)^k\Big),&t\in F_q\setminus \{0\}
\end{cases}
\end{split}
\end{equation}
\end{lemma}
\begin{IEEEproof}
See Lemma 2 in \cite{2019Chen}.
\end{IEEEproof}

The following lemma generalizes (4) in \cite{2019Chen}. It reveals that, for non-zero elements $g_1,g_2,\cdots,g_k$, the random variable $g_1X_1+g_2X_2+\cdots+g_kX_k$ has same distribution as the random variable $X_1+X_2+\cdots+X_k$, whether the random variables $X_1,X_2,\cdots,X_k$ are independent or not.
\begin{lemma}\label{lemma2}
Let $X_1,X_2,\cdots,X_k$ be dependent random variables over $F_q$ and have a marginal distribution (\ref{equation1}) with different parameters $p_1,p_2,\cdots,p_k$. Let $g_1,g_2,\cdots,g_k$ be non-zero elements from $F_q$. Then
\begin{equation}
\begin{split}
&Pr\{g_1X_1+g_2X_2+\cdots+g_kX_k=t\}\\
=&Pr\{X_1+X_2+\cdots+X_k=t\}
\end{split}
\end{equation}
\end{lemma}
\begin{IEEEproof}
See the appendix.
\end{IEEEproof}

In general, it is not easy to obtain the exact distribution of each entry of the inverse matrix of a random matrix. The following lemma provides an approximation to it based on the general linear group, and all approximations in this paper are related to this approximation.

\begin{lemma}\label{lemma3}
Let $H$ be a random $i\times i$ matrix over $F_q$, whose entries are i.i.d. with (\ref{equation1}), and assume that $H$ is nonsingular. Then each entry of the inverse matrix $H^{-1}$ of the matrix $H$ can be approximated as i.i.d. with (\ref{equation1}).
\end{lemma}
\begin{IEEEproof}
See the appendix.
\end{IEEEproof}

In the following lemma, by presenting the explicit structure of the RREF of a full rank matrix, we derive a novel approximation to the probability $p(i,n)$.

\begin{lemma}\label{lemma4}
Let $H$ be a random $(i+1)\times n$ matrix over $F_q$, whose entries are i.i.d. with (\ref{equation1}), $0\le i \le n-1$, and assume that the first $i$ rows of $H$ are linearly independent. Then the probability of $H$ being not full rank can be approximated as
\begin{equation}\label{equation4}
\begin{split}
p(i,n)\cong&\sum_{k=0}^{i}\binom{i}{k}(1-p_0)^{k}(p_0)^{i-k}\\
&\Bigg(\frac{1}{q}+(p_0-\frac{1}{q})\Big(1-\frac{q(1-p_0)(1-p_{\Delta,k})}{q-1}\Big)^{i}\Bigg)^{n-i},
\end{split}
\end{equation}
where $p_{\Delta,k}\cong \beta_{k}(p_0,t=0)=\frac{1}{q}\Big(1+(q-1)\big(1-\frac{q(1-p_0)}{q-1}\big)^{k}\Big)$.
\end{lemma}
\begin{IEEEproof}
Let $h_k$ be the $k$-th row of $H$, $k=1,2,\cdots,i+1$, and $A=(h_1,h_2,\cdots,h_i)^{T}$ be a $i\times n$ matrix composed of the first $i$ rows of $H$. Then $H$ is not full rank if and only if $h_1,h_2,\cdots,h_{i+1}$ are linearly dependent. Since $h_1,h_2,\cdots,h_{i}$ are linearly independent, $h_{i+1}$ can be uniquely expressed linearly by $h_1,h_2,\cdots,h_{i}$. In other words, there is a vector $c=(c_1,c_2,\cdots,c_i) \in F_q^{1\times i}$ such that
\begin{equation*}
h_{i+1}^{T}=c_1h_{1}^{T}+c_2h_{2}^{T}+\cdots+c_ih_{i}^{T},
\end{equation*}
or in matrix form,
\begin{equation}\label{equation5}
h_{i+1}=cA.
\end{equation}
It is easy to verify that the set $V=\{c_1h_1+c_2h_2+\cdots+c_ih_i~|~c_1,c_2,\cdots,c_i \in F_q\}$ is a vector space over $F_q$, called the vector space generated by $h_1,h_2,\cdots,h_{i}$, and $h_1,h_2,\cdots,h_{i}$ is a basis of vector space $V$. From a perspective of vector space, $H$ is not full rank if and only if $h_{i+1}$ is contained in the subspace generated by $h_1,h_2,\cdots,h_{i}$. Since the basis of vector space is not unique and the coordinates of a vector under the natural basis are its entries, it is natural to reduce the basis $h_1,h_2,\cdots,h_{i}$ to a basis like the natural basis.

We first consider the case that the first $i$ columns of $A$ are linearly independent. Thus, they are a maximal linearly independent set of the columns of $A$, and therefore the last $n-i$ columns of $A$ can be expressed as linear combinations of this maximal linearly independent set. Let $A_1$ be a $i\times i$ matrix composed of the first $i$ columns of $A$ and $A_2$ be a $i\times (n-i)$ matrix composed of the last $n-i$ columns of $A$, then $rk(A_1)=i$ and $A_2=A_1A'_2$, where $A'_2 \in F_q^{i\times (n-i)}$ is the coefficient matrix that $A_2$ is linearly expressed by $A_1$. Therefore, we have
\begin{equation*}
A=(A_1,A_2)=(A_1,A_1A'_2)=A_1(I, A'_2),
\end{equation*}
or
\begin{equation*}
A_1^{-1}A=(I, A'_2).
\end{equation*}
Let $A'=(I, A'_2)$. From linear algebra, $A'$ is actually the RREF of the matrix $A$ (exist and is unique). According to the basis change formula, the rows of $A'$ are also a basis of $V$, and then (\ref{equation5}) can be also written as
\begin{equation}\label{equation6}
h_{i+1}=gA',
\end{equation}
where $g=(g_1,g_2,\cdots,g_i) \in F_q^{1\times i}$.

If the first $i$ columns of $A$ are not linearly independent, we can interchange the columns of $A$ such that the first $i$ columns of $A$ are linearly independent. This is equivalent to $A$ is multiplied by an $n\times n$ invertible matrix $Q$ on the right such that the first $i$ columns of $A$ are linearly independent, i.e., $AQ=B=(B_1,B_2)$, where $B_1$ is a $i\times i$ matrix composed of a maximal linearly independent set of the columns of $A$, $B_2$ is a $i\times (n-i)$ matrix composed of the columns of $A$ except for the maximal linearly independent set. It is worth noting that $B$ is different from $A$ only at the order of the columns. Then $rk(B_1)=i$ and $B_2=B_1B'_2$, where $B'_2\in F_q^{i\times (n-i)}$.
Therefore, we have
\begin{equation*}
AQ=B=(B_1,B_2)=(B_1,B_1B'_2)=B_1(I,B'_2),
\end{equation*}
or
\begin{equation*}
B_1^{-1}A=(I,B'_2)Q^{-1}.
\end{equation*}
Then,
\begin{equation*}
h_{i+1}=g(I,B'_2)Q^{-1},
\end{equation*}
or
\begin{equation*}
h_{i+1}Q=g(I,B'_2).
\end{equation*}
Since each entry of $h_{i+1}$ is i.i.d. with (\ref{equation1}) and $Q$ is the product of elementary matrices that interchanging two columns, each entry of $h_{i+1}Q$ is still i.i.d. with (\ref{equation1}), and therefore $h_{i+1}Q$ is essentially same to $h_{i+1}$.

The matrix $A'$ is given by
\begin{equation}\label{equation7}
A'=(I, A'_2)=
\begin{pmatrix}
  1 & \cdots & 0 & a'_{1,i+1} & \cdots & a'_{1,n} \\
  \vdots & \ddots & \vdots & \vdots & \ddots & \vdots \\
  0 & \cdots & 1 & a'_{i,i+1} & \cdots & a'_{i,n}
\end{pmatrix}.
\end{equation}
Substituting (\ref{equation7}) into (\ref{equation6}), the column $h_{i+1}^{T}$ is given by
\begin{equation*}
\begin{split}
  h_{i+1}^{T}=&(A')^{T}g^{T}\\
  =&g_1\begin{pmatrix}
         1 \\
         0 \\
         \vdots \\
         0 \\
         a'_{1,i+1} \\
         \vdots \\
         a'_{1,n}
       \end{pmatrix}
       +g_2\begin{pmatrix}
         0 \\
         1 \\
         \vdots \\
         0 \\
         a'_{2,i+1} \\
         \vdots \\
         a'_{2,n}
       \end{pmatrix}
       +\cdots
       +g_i\begin{pmatrix}
         0 \\
         0 \\
         \vdots \\
         1 \\
         a'_{i,i+1} \\
         \vdots \\
         a'_{i,n}
       \end{pmatrix}  \\
       =&\begin{pmatrix}
          g_1 \\
          g_2 \\
          \vdots \\
          g_i \\
          g_1a'_{1,i+1}+g_2a'_{2,i+1}+\cdots+g_ia'_{i,i+1} \\
          \vdots \\
          g_1a'_{1,n}+g_2a'_{2,n}+\cdots+g_ia'_{i,n}
        \end{pmatrix}.
\end{split}
\end{equation*}
From above equation, $h_{i+1}$ is contained in the subspace generated by $h_1,h_2,\cdots,h_{i}$ if and only if the first $i$ entries of $h_{i+1}$ are arbitrary and the last $n-i$ entries are uniquely determined by the first $i$ entries. Therefore the left work is to calculate the joint distribution of the last $n-i$ entries of $h_{i+1}$.

Since $A'_2=A_1^{-1}A_2$ and the entries of $A_1^{-1}$ are not independent, the entries $a'_{k,l}$ of $A'_2$ are not independent especially for the entries in the same column or row, where $k=1,\cdots,i$ and $l=i+1,\cdots,n$. Therefore, the entries $a'_{k,l}$ need to be further decomposed into independent or approximately independent components.

Let $(A_1)^{-1}=\Delta=(\Delta_{k,l})$, where $k,l=1,2,\cdots,i$. Then the $(i+1)$-th and the $n$-th entry of $h_{i+1}$ are
\begin{equation*}
\begin{split}
&g_1a'_{1,i+1}+g_2a'_{2,i+1}+\cdots+g_ia'_{i,i+1}\\
=&g_1(\Delta_{1,1}a_{1,i+1}+\Delta_{1,2}a_{2,i+1}+\cdots+\Delta_{1,i}a_{i,i+1})\\
&+g_2(\Delta_{2,1}a_{1,i+1}+\Delta_{2,2}a_{2,i+1}+\cdots+\Delta_{2,i}a_{i,i+1})\\
&+\cdots\\
&+g_i(\Delta_{i,1}a_{1,i+1}+\Delta_{i,2}a_{2,i+1}+\cdots+\Delta_{i,i}a_{i,i+1})
\end{split}
\end{equation*}
and
\begin{equation*}
\begin{split}
&g_1a'_{1,n}+g_2a'_{2,n}+\cdots+g_ia'_{i,n}\\
=&g_1(\Delta_{1,1}a_{1,n}+\Delta_{1,2}a_{2,n}+\cdots+\Delta_{1,i}a_{i,n})\\
&+g_2(\Delta_{2,1}a_{1,n}+\Delta_{2,2}a_{2,n}+\cdots+\Delta_{2,i}a_{i,n})\\
&+\cdots\\
&+g_i(\Delta_{i,1}a_{1,n}+\Delta_{i,2}a_{2,n}+\cdots+\Delta_{i,i}a_{i,n})
\end{split}
\end{equation*}
Note that the last $n-i$ entries of $h_{i+1}$ have common components $g_1,g_2,\cdots,g_i$ and $\Delta_{k,l}$.
Consider that $g_1,g_2,\cdots,g_i$ are independent, it is easy to obtain their joint distribution. Let $g=(g_1,g_2,\cdots,g_i)$ have exactly $k$ non-zero entries. Without loss of generality, we assume that the first $k$ entries of $g$ are non-zero. Let $\Delta_{l}^{(k)}=\Delta_{1,l}+\Delta_{2,l}+\cdots+\Delta_{k,l}$ be the sum of the first $k$ entries of the $l$-th column of $\Delta$, where $k,l=1,2,\cdots,i$. Then, according to Lemma \ref{lemma3} and Lemma \ref{lemma1}, $p_{\Delta,k}=Pr\{\Delta_{l}^{(k)}=0\}\cong \beta_{k}(p_0,t=0)$.
Since $A'_2=A_1^{-1}A_2$, according to Lemma \ref{lemma3} and Lemma \ref{lemma1}, $a'_{k,l}$ is approximately distributed with (\ref{equation1}), but the parameter is not $p_0$.
Therefore, according to Lemma \ref{lemma2} and Lemma \ref{lemma1},
\begin{align*}
&Pr\{g_1a'_{1,i+1}+g_2a'_{2,i+1}+\cdots+g_ia'_{i,i+1}=t\}\\
=&Pr\{g_1a'_{1,i+1}+g_2a'_{2,i+1}+\cdots+g_ka'_{k,i+1}=t\}\\
\cong&Pr\{a'_{1,i+1}+a'_{2,i+1}+\cdots+a'_{k,i+1}=t\}\\
=&Pr\{\Delta_{1}^{(k)}a_{1,i+1}+\Delta_{2}^{(k)}a_{2,i+1}+\cdots+\Delta_{i}^{(k)}a_{i,i+1}=t\}\\
\cong&\beta_{i}(p_0+(1-p_0)p_{\Delta,k}, t).
\end{align*}

Therefore, according to total probability theorem, the probability of $h_{i+1}$ being contained in the subspace generated by $h_1,h_2,\cdots,h_{i}$ is given by
\begin{align}\label{equation17}
&p(i,n)\nonumber\\
\cong& \sum_{k=0}^{i}\binom{i}{k}(1-p_0)^{k}(p_0)^{i-k}\nonumber\\
&\Big(p_0\beta_{i}(p_0+(1-p_0)p_{\Delta,k}, t=0)\nonumber\\
&+(1-p_0)\beta_{i}(p_0+(1-p_0)p_{\Delta,k}, t\neq 0)\Big)^{n-i}\nonumber\\
=&\sum_{k=0}^{i}\binom{i}{k}(1-p_0)^{k}(p_0)^{i-k}\nonumber\\
&\Bigg(p_0\frac{1}{q}\Big(1+(q-1)\Big(1-\frac{q(1-p_0)(1-p_{\Delta,k})}{q-1}\Big)^{i}\Big)\nonumber\\
&+(1-p_0)\frac{1}{q}\Big(1+(-1)\Big(1-\frac{q(1-p_0)(1-p_{\Delta,k})}{q-1}\Big)^{i}\Big)\Bigg)^{n-i}\nonumber\\
=&\sum_{k=0}^{i}\binom{i}{k}(1-p_0)^{k}(p_0)^{i-k}\nonumber\\
&\Bigg(\frac{1}{q}+(p_0-\frac{1}{q})\Big(1-\frac{q(1-p_0)(1-p_{\Delta,k})}{q-1}\Big)^{i}\Bigg)^{n-i}.
\end{align}
The lemma follows.
\end{IEEEproof}

\textit{Remark:} It is well-known that the probability of a vector being contained in the subspace generated by $i$ linearly independent vectors is $(1/q)^{n-i}$ for RLNC scheme \cite{2000Cooper,2011Trullols-Cruces}. By substituting $p_0=1/q$ into (\ref{equation4}), $p(i,n)=(1/q)^{n-i}$ which shows that our derived approximation can degenerate into RLNC case.

\subsection{The Probability of Sparse Random Matrices Being Full Rank}
In the following theorem, we derive an exact expression for the probability $P(n,m)$ as a function of $p(i,n)$ based on the product theorem, and gives an approximation to $P(n,m)$.
\newtheorem{theorem}{Theorem}
\begin{theorem}\label{theorem1}
Let $M$ be a random $n\times m$ matrix over $F_q$, whose entries are i.i.d. with (\ref{equation1}), $n\le m$. Then the probability of $M$ being full rank can be approximated as
\begin{align}
&P(n,m)\nonumber\\
=&\prod_{i=0}^{n-1}\big(1-p(i,m)\big)\nonumber\\
\label{equation9}\cong&\prod_{i=0}^{n-1}\Bigg(1-\sum_{k=0}^{i}\binom{i}{k}(1-p_0)^{k}(p_0)^{i-k}\nonumber\\
&\Big(\frac{1}{q}+(p_0-\frac{1}{q})\Big(1-\frac{q(1-p_0)(1-p_{\Delta,k})}{q-1}\Big)^{i}\Big)^{m-i}\Bigg),
\end{align}
where $p_{\Delta,k}\cong \beta_{k}(p_0,t=0)=\frac{1}{q}\Big(1+(q-1)\big(1-\frac{q(1-p_0)}{q-1}\big)^{k}\Big)$.
\end{theorem}
\begin{IEEEproof}
Let $M_{i}$ be a $i\times m$ matrix composed of the first $i$ rows of $M$. Obviously, $Pr\{rk(M_{0})=0\}=1$.
Since $M$ has rank $n$ if and only if the first $i$ rows of $M$ are linearly independent, $0\le i\le n$,
\begin{equation*}
\begin{split}
&Pr\{rk(M)=n\}\\
=&Pr\{rk(M_{n})=n, rk(M_{n-1})=n-1,\cdots,rk(M_{0})=0\}.
\end{split}
\end{equation*}
According to the product theorem,
\begin{align*}
&Pr\{rk(M)=n\}\\
=&Pr\{rk(M_{n})=n~|rk(M_{n-1})=n-1,\cdots,rk(M_{0})=0\}\\
&Pr\{rk(M_{n-1})=n-1~|rk(M_{n-2})=n-2,\cdots,rk(M_{0})=0\}\\
&\cdots\\
&Pr\{rk(M_{1})=1~|rk(M_{0})=0\}\\
&Pr\{rk(M_{0})=0\}\\
=&Pr\{rk(M_{n})=n~|rk(M_{n-1})=n-1\}\\
&Pr\{rk(M_{n-1})=n-1~|rk(M_{n-2})=n-2\}\\
&\cdots\\
&Pr\{rk(M_{1})=1~|rk(M_{0})=0\}\\
&Pr\{rk(M_{0})=0\}\\
=&\prod_{i=0}^{n-1}Pr\{rk(M_{i+1})=i+1~|rk(M_{i})=i\}\\
=&\prod_{i=0}^{n-1}\big(1-p(i,m)\big)
\end{align*}
Then, by substituting (\ref{equation4}) into above equation, (\ref{equation9}) holds.
\end{IEEEproof}

\subsection{Complexity Analysis And Comparison}
In this subsection, the complexity of our proposed approximation and other state-of-the-art approximations provided in \cite{2018Brown,2019Chen,2019Sehat} are derived and compared.

\begin{theorem}\label{theorem2}
The complexity of (\ref{equation9}) is $O(n^2)$.
\end{theorem}
\begin{IEEEproof}
The complexity of the inner sum in (\ref{equation9}) is $O(n)$. Therefore, the complexity of (\ref{equation9}) is $O(n\times n)$.
\end{IEEEproof}

Before deriving the complexity of other state-of-the-art approximations, we first present their expressions.

In \cite{2018Brown}, $P(n,m)$ can be approximated as follows:
\begin{equation}\label{equation-Brown}
\begin{split}
P(n,m)\cong (1-p_{0}^{m})^{n}exp\Big(-\sum_{l=2}^{n}\binom{n}{l}\frac{\tilde{\pi}_{l}}{(1-p_0^{m})^{l}}\Big),
\end{split}
\end{equation}
where
$\tilde{\pi}_{l}=\rho_{l}-\sum_{s=1}^{l-1}\binom{l-1}{s}\rho_{s}\tilde{\pi}_{l-s}$,\newline
$\rho_{l}=\Bigg(\frac{1}{q}\Big(1+(q-1)\big(1-\frac{q(1-p_0)}{q-1}\big)^{l}\Big)\Bigg)^{m}$
and $\tilde{\pi}_{1}=\rho_{1}$.

In \cite{2019Chen}, $P(n,m)$ can be approximated as follows:
\begin{equation}\label{equation-Chen}
P(n,m)\cong (1-p_0^{m})^{n}\prod_{i=0}^{n-1}(1-s^{m-i}),
\end{equation}
where
$s=\frac{1}{q}+(p_0-\frac{1}{q})\Big(1-\frac{q(1-p_0)(1-p_{a'})}{q-1}\Big)^{n}$,\newline
$p_{a'}\cong \frac{1}{q}\Big(1+(q-1)(1-\frac{q(1-p_0)^2}{q-1})^{n}\Big)$.

In \cite{2019Sehat}, $P(n,m)$ can be approximated as follows:
\begin{equation}\label{equation-Sehat}
P(n,m)\cong \prod_{k=1}^{n}(1-b_k)^{n_k},
\end{equation}
where
$b_k=\Big(\frac{q-1}{q}(1-\frac{q(1-p_0)}{q-1})^{k}+\frac{1}{q}\Big)^m$,
$n_k=\binom{n}{k}(q-1)^{k}$.

\begin{theorem}\label{theorem3}
The complexity of (\ref{equation-Brown}), (\ref{equation-Chen}) and (\ref{equation-Sehat}) are $O(n^3)$, $O(n)$ and $O(n)$, respectively.
\end{theorem}
\begin{IEEEproof}
The most computationally intensive part of  (\ref{equation-Brown}) is $\tilde{\pi}_{l}$. We note that by running $\tilde{\pi}_{l}$, the values of $\tilde{\pi}_{1},\cdots,\tilde{\pi}_{l-1}$ will also be obtained. Thus, the complexity of $\tilde{\pi}_{l}$ is $O(n^2)$ and the complexity of (\ref{equation-Brown}) is $O(n\times n^2)$. The complexity of (\ref{equation-Chen}) and (\ref{equation-Sehat}) are straightforward.
\end{IEEEproof}

\textit{Remark:} From Theorem \ref{theorem2} and Theorem \ref{theorem3}, we can observe that, the complexity of our proposed approximation (\ref{equation9}) is lower than (\ref{equation-Brown}), but is higher than (\ref{equation-Chen}) and (\ref{equation-Sehat}). However, as shown in Section IV, the accuracy of our proposed approximation (\ref{equation9}) is higher than (\ref{equation-Chen}) for almost all cases and (\ref{equation-Sehat}) for all cases.

\section{Simulation Results}
\label{section4}
In this section, we validate the accuracy of our derived approximation by Monte Carlo simulations and compare it with the state-of-the-art approximations in literatures.
In order to thoroughly assess different approximations, we set $n=\{8,32,128\},q=\{2,256\},p_0=\{1/q,0.8,0.9,0.96\}$.
In addition, in order to ensure statistical tightness of results, all simulation results are the average of $10^{10},10^{9},10^{7}$ independent runs for $n=8,32,128$, respectively. The deviation between the approximation and simulation result is evaluated in terms of Mean Squared Error (MSE).

\begin{figure*}
\centering
  \subfigure[$q=2,n=8$]{
  \includegraphics[width=0.5\textwidth]{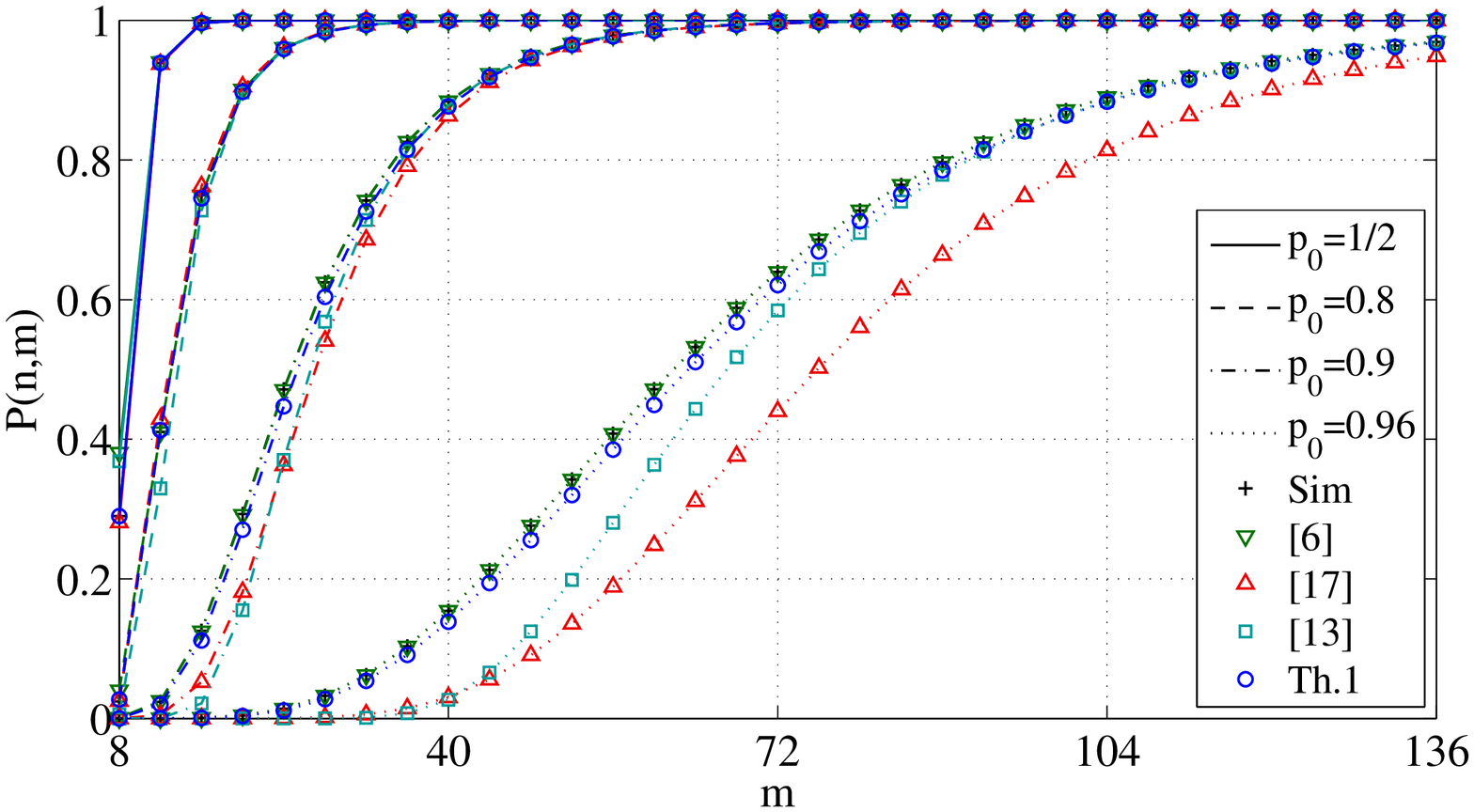}}
  \hspace{-1cm}
  \subfigure[$q=256,n=8$]{
  \includegraphics[width=0.5\textwidth]{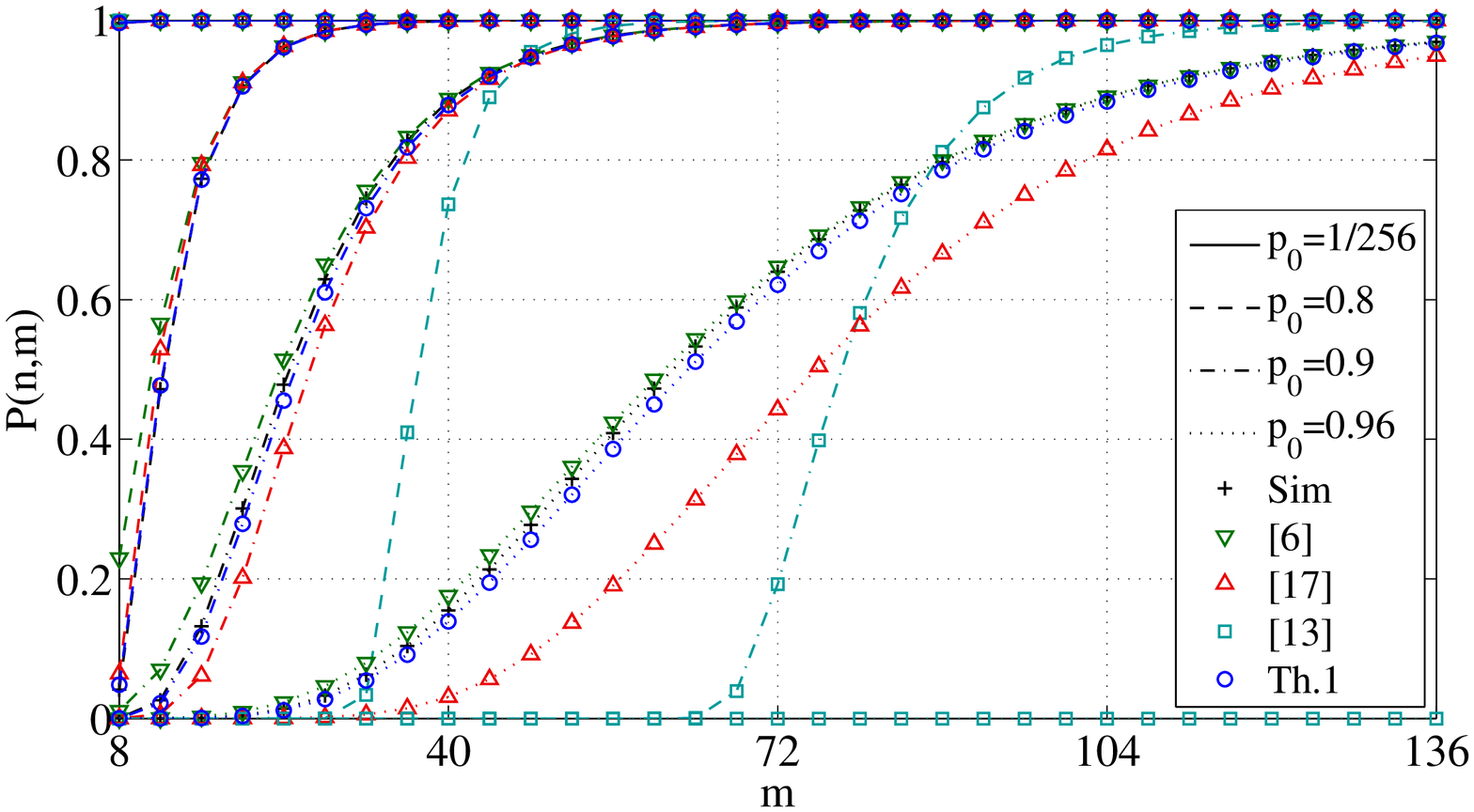}}

  \subfigure[$q=2,n=32$]{
  \label{fig1c}
  \includegraphics[width=0.5\textwidth]{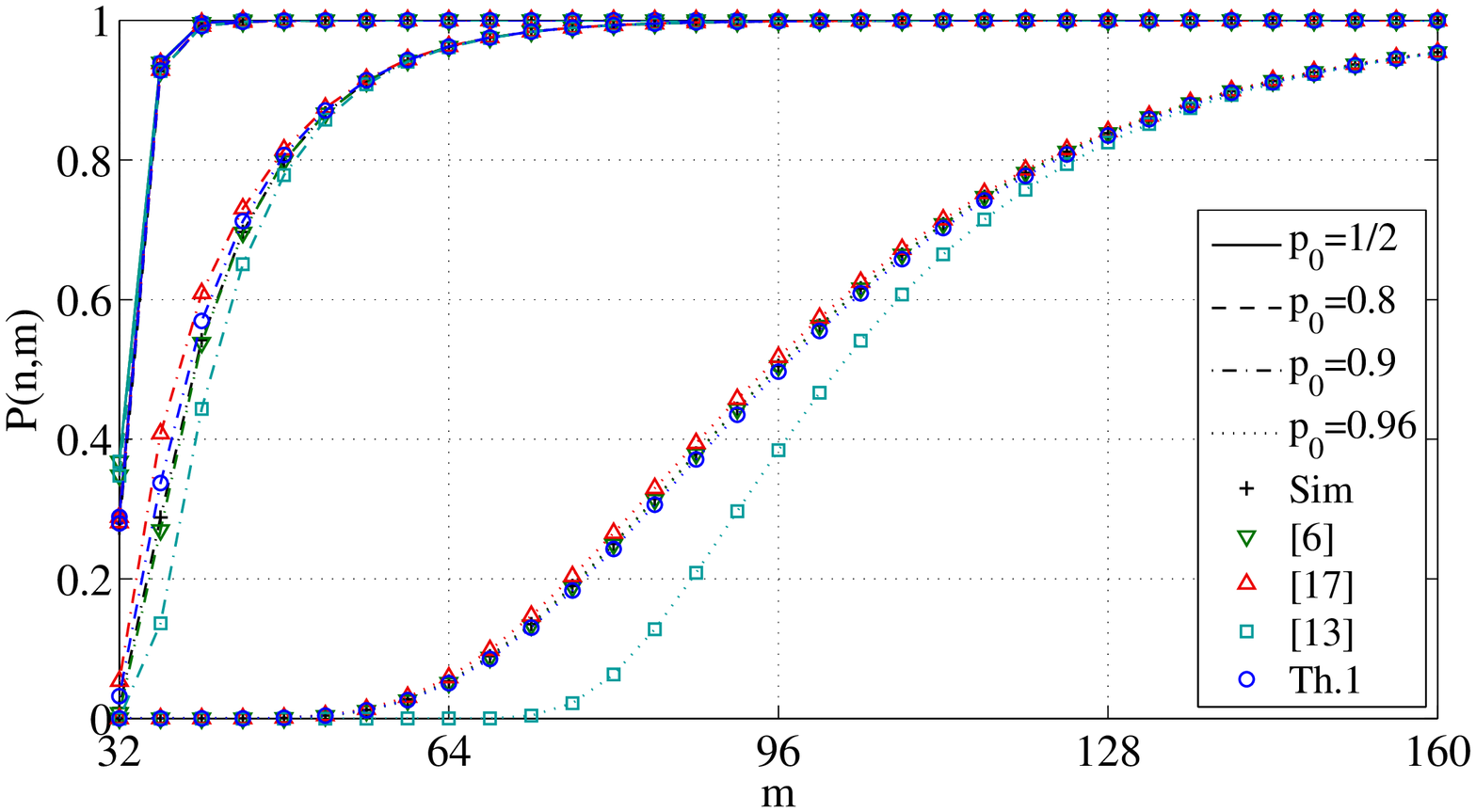}}
  \hspace{-1cm}
  \subfigure[$q=256,n=32$]{
  \label{fig1d}
  \includegraphics[width=0.5\textwidth]{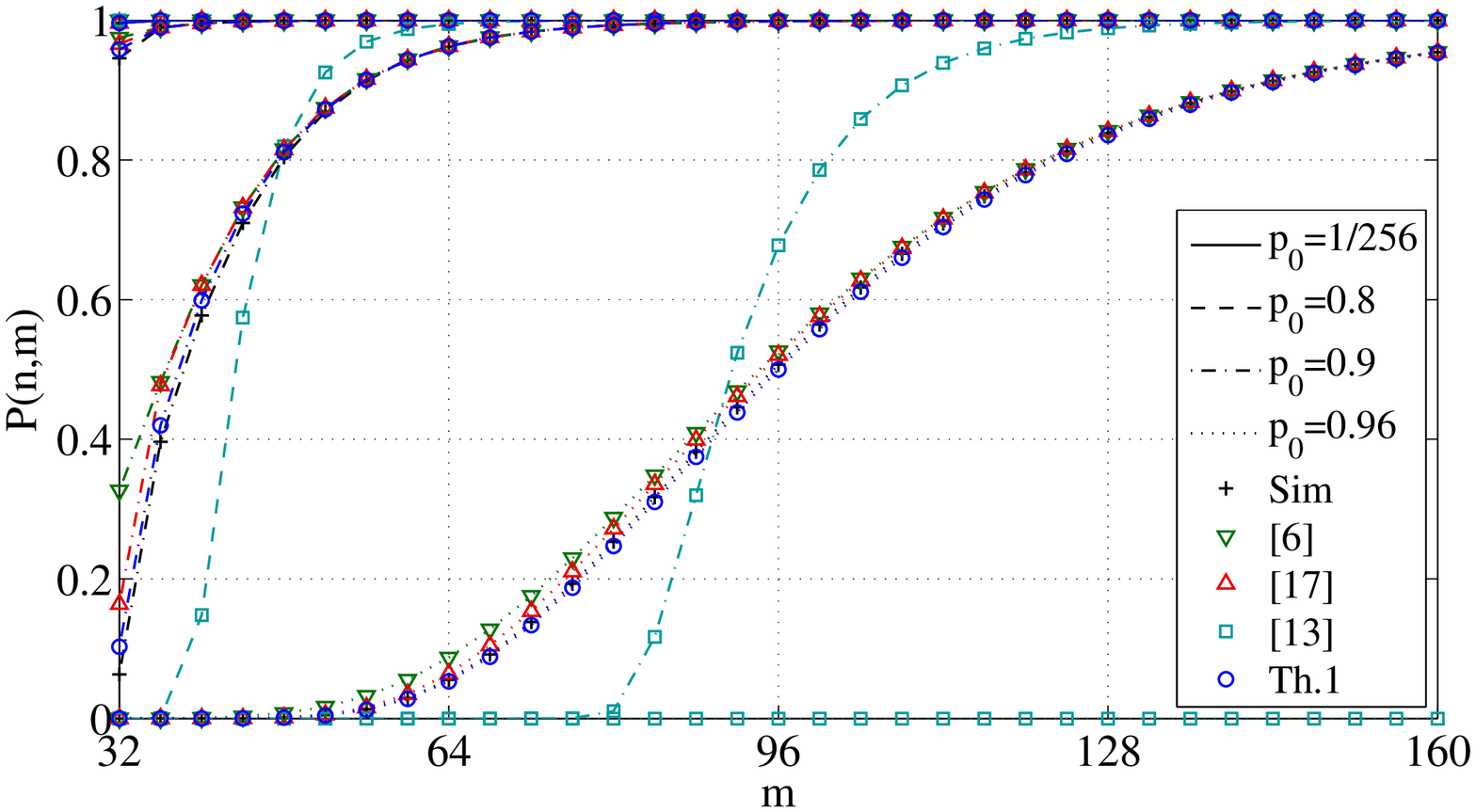}}

  \subfigure[$q=2,n=128$]{
  \label{fig1e}
  \includegraphics[width=0.5\textwidth]{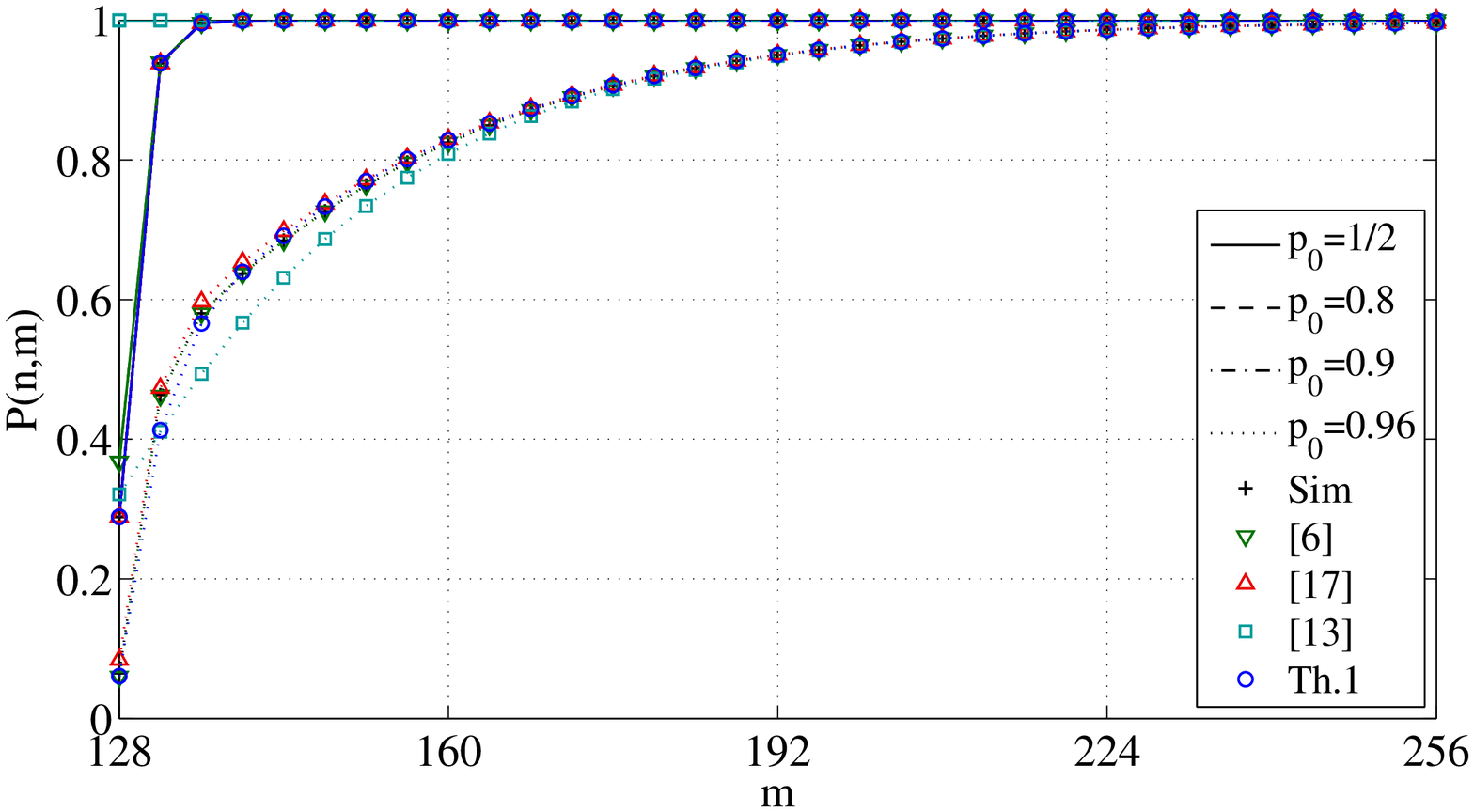}}
  \hspace{-1cm}
  \subfigure[$q=256,n=128$]{
  \label{fig1f}
  \includegraphics[width=0.5\textwidth]{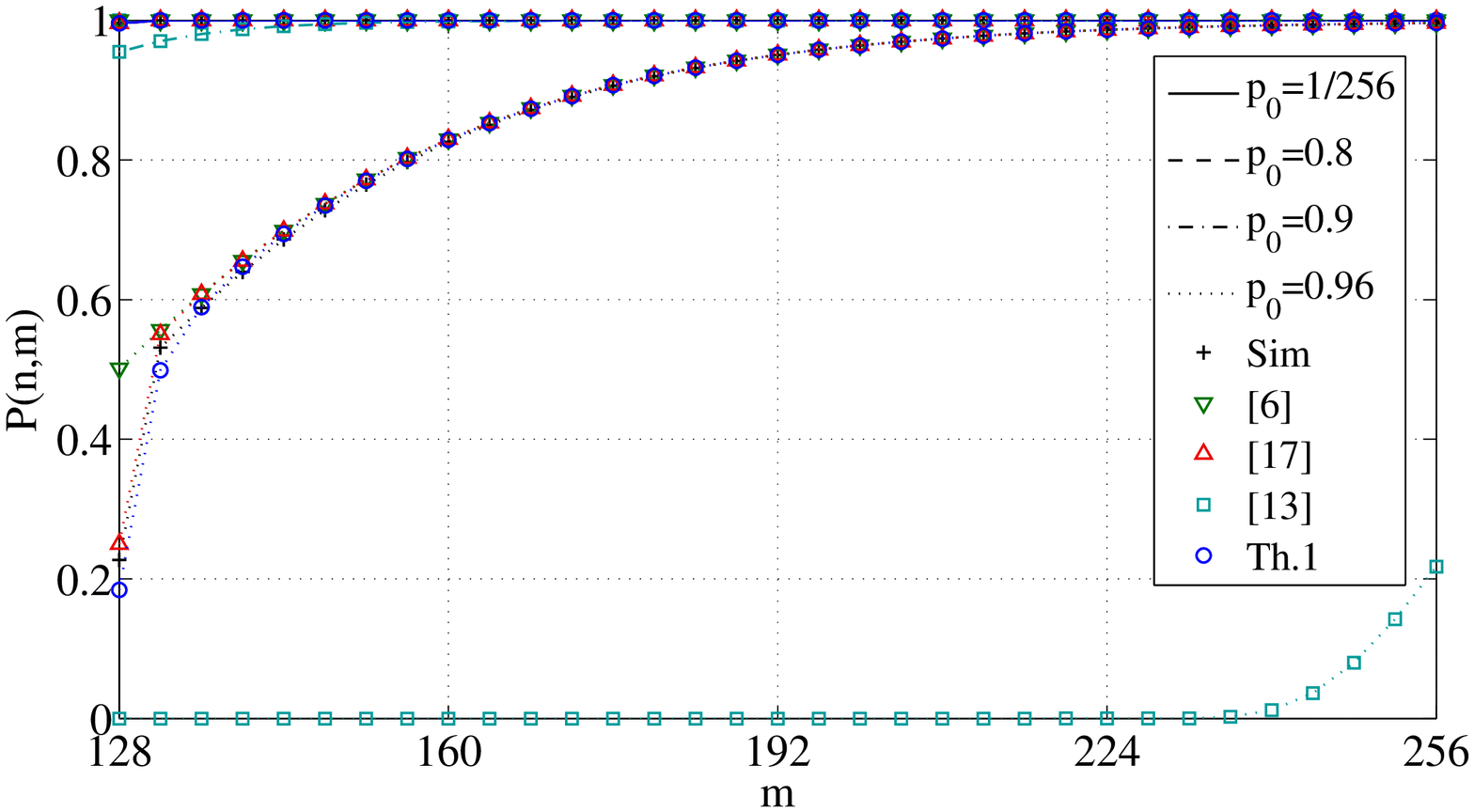}}

  \caption{The probability $P(n,m)$ as a function of $m$, for $q=\{2,256\}$, $n=\{8,32,128\}$ and $p_0=\{1/q,0.8,0.9,0.96\}$.}
  \label{fig1}
\end{figure*}

\definecolor{myred}{rgb}{0.8,0,0}
\begin{table*}[!t]
\renewcommand{\arraystretch}{1.3}
\centering
\caption{The MSE of \cite{2018Brown},\cite{2019Chen},\cite{2019Sehat} and Th.1, for $P(n,m)$, in Different Configurations}
\label{table1}
\begin{tabular}{|c|c|c|c|c|c|}
\hline
\multicolumn{2}{|c|}{~} & $p_0=1/q$ & $p_0=0.8$ & $p_0=0.9$ & $p_0=0.96$ \\\hline
\multirow{3}*{$q=2,n=8$} & \cite{2018Brown} & $\color{myred}{\mathbf{7.36\cdot10^{-5}}}$ & $3.34\cdot10^{-6}$ & $\color{blue}{\mathbf{1.51\cdot10^{-7}}}$ & $\color{blue}{\mathbf{1.09\cdot10^{-8}}}$\\
~ & \cite{2019Chen} & $1.65\cdot10^{-6}$ & $1.56\cdot10^{-5}$ & $1.30\cdot10^{-3}$ & $\color{myred}{\mathbf{1.60\cdot10^{-2}}}$ \\
~ & \cite{2019Sehat} & $5.50\cdot 10^{-5}$ & $\color{myred}{\mathbf{2.58\cdot 10^{-4}}}$ & $\color{myred}{\mathbf{1.38\cdot 10^{-3}}}$ & $4.45\cdot 10^{-3}$ \\
~ & Th.1 & $\color{blue}{\mathbf{6.63\cdot10^{-13}}}$ & $\color{blue}{\mathbf{1.97\cdot10^{-6}}}$ & $6.72\cdot10^{-5}$ & $1.63\cdot10^{-4}$ \\\hline

\multirow{3}*{$q=2,n=32$} & \cite{2018Brown} & $\color{myred}{\mathbf{5.57\cdot10^{-5}}}$ & $\color{myred}{\mathbf{5.02\cdot10^{-5}}}$ & $\color{blue}{\mathbf{1.80\cdot10^{-5}}}$ & $\color{blue}{\mathbf{6.89\cdot10^{-7}}}$\\
~ & \cite{2019Chen} & $\color{blue}{\mathbf{2.55\cdot10^{-12}}}$ & $1.41\cdot10^{-6}$ & $7.36\cdot10^{-4}$ & $5.90\cdot10^{-5}$ \\
~ & \cite{2019Sehat} & $5.56\cdot 10^{-5}$ & $4.95\cdot 10^{-5}$ & $\color{myred}{\mathbf{1.08\cdot 10^{-3}}}$ & $\color{myred}{\mathbf{6.65\cdot 10^{-3}}}$ \\
~ & Th.1 & $\color{blue}{\mathbf{2.55\cdot10^{-12}}}$ & $\color{blue}{\mathbf{8.23\cdot10^{-7}}}$ & $1.30\cdot10^{-4}$ & $2.17\cdot10^{-5}$ \\\hline

\multirow{3}*{$q=2,n=128$} & \cite{2018Brown} & $5.59\cdot10^{-5}$ & $5.57\cdot10^{-5}$ & $5.55\cdot10^{-5}$ & $\textcolor{blue}{\mathbf{1.50\cdot10^{-6}}}$\\
~ & \cite{2019Chen} & $\color{blue}{\mathbf{5.84\cdot10^{-10}}}$ & $\color{blue}{\mathbf{1.58\cdot10^{-10}}}$ & $\color{blue}{\mathbf{1.33\cdot10^{-10}}}$ & $4.60\cdot10^{-5}$ \\
~ & \cite{2019Sehat} & $\color{myred}{\mathbf{5.87\cdot 10^{-3}}}$ & $\color{myred}{\mathbf{5.87\cdot 10^{-3}}}$ & $\color{myred}{\mathbf{5.86\cdot 10^{-3}}}$ & $\color{myred}{\mathbf{1.41\cdot 10^{-3}}}$ \\
~ & Th.1 & $\color{blue}{\mathbf{5.84\cdot10^{-10}}}$ & $\color{blue}{\mathbf{1.58\cdot10^{-10}}}$ & $\color{blue}{\mathbf{1.33\cdot10^{-10}}}$ & $9.25\cdot10^{-5}$ \\\hline

\multirow{3}*{$q=256,n=8$} & \cite{2018Brown} & $\color{myred}{\mathbf{1.19\cdot10^{-7}}}$ & $1.00\cdot10^{-3}$ & $3.29\cdot10^{-4}$ & $\color{blue}{\mathbf{1.01\cdot10^{-4}}}$\\
~ & \cite{2019Chen} & $\color{blue}{\mathbf{1.00\cdot10^{-15}}}$ & $1.50\cdot10^{-4}$ & $9.51\cdot10^{-4}$ & $1.57\cdot10^{-2}$ \\
~ & \cite{2019Sehat} & $\color{myred}{\mathbf{1.19\cdot 10^{-7}}}$ & $\color{myred}{\mathbf{1.51\cdot 10^{-1}}}$ & $\color{myred}{\mathbf{3.28\cdot 10^{-1}}}$ & $\color{myred}{\mathbf{4.23\cdot 10^{-1}}}$ \\
~ & Th.1 & $\color{blue}{\mathbf{1.00\cdot10^{-15}}}$ & $\color{blue}{\mathbf{3.56\cdot10^{-6}}}$ & $\color{blue}{\mathbf{6.00\cdot10^{-5}}}$ & $1.63\cdot10^{-4}$ \\\hline

\multirow{3}*{$q=256,n=32$} & \cite{2018Brown} & $\color{myred}{\mathbf{1.19\cdot10^{-7}}}$ & $6.63\cdot10^{-6}$ & $1.30\cdot10^{-3}$ & $3.32\cdot10^{-4}$\\
~ & \cite{2019Chen} & $\color{blue}{\mathbf{1.55\cdot10^{-15}}}$ & $3.23\cdot10^{-6}$ & $5.60\cdot10^{-4}$ & $8.19\cdot10^{-5}$ \\
~ & \cite{2019Sehat} & $\color{myred}{\mathbf{1.19\cdot 10^{-7}}}$ & $\color{myred}{\mathbf{7.77\cdot 10^{-2}}}$ & $\color{myred}{\mathbf{3.27\cdot 10^{-1}}}$ & $\color{myred}{\mathbf{3.53\cdot 10^{-1}}}$ \\
~ & Th.1 & $\color{blue}{\mathbf{1.55\cdot10^{-15}}}$ & $\color{blue}{\mathbf{1.20\cdot10^{-6}}}$ & $\color{blue}{\mathbf{6.89\cdot10^{-5}}}$ & $\color{blue}{\mathbf{1.66\cdot10^{-5}}}$ \\\hline

\multirow{3}*{$q=256,n=128$} & \cite{2018Brown} & $\color{myred}{\mathbf{1.19\cdot10^{-7}}}$ & $\color{myred}{\mathbf{1.20\cdot10^{-7}}}$ & $1.32\cdot10^{-7}$ & $7.43\cdot10^{-4}$\\
~ & \cite{2019Chen} & $\color{blue}{\mathbf{4.91\cdot10^{-14}}}$ & $\color{blue}{\mathbf{2.06\cdot10^{-12}}}$ & $1.77\cdot10^{-10}$ & $\color{blue}{\mathbf{4.51\cdot10^{-5}}}$ \\
~ & \cite{2019Sehat} & $\color{myred}{\mathbf{1.19\cdot 10^{-7}}}$ & $\color{myred}{\mathbf{1.20\cdot 10^{-7}}}$ & $\color{myred}{\mathbf{8.05\cdot 10^{-5}}}$ & $\color{myred}{\mathbf{7.75\cdot 10^{-1}}}$ \\
~ & Th.1 & $\color{blue}{\mathbf{4.91\cdot10^{-14}}}$ & $\color{blue}{\mathbf{2.06\cdot10^{-12}}}$ & $\color{blue}{\mathbf{1.31\cdot10^{-11}}}$ & $1.56\cdot10^{-4}$ \\\hline

\end{tabular}
\end{table*}

Fig. \ref{fig1} shows the probability of a decoding matrix being full rank $P(n,m)$ as a function of the received coded packets $m$. The corresponding MSEs of the approximations are provided in Table \ref{table1}.
As can be seen from the figure, the difference between our proposed approximation and simulation results is negligible. The maximum MSE of our proposed approximation is occurred at $q=\{2,256\},n=8,p_0=0.96$ and is equal to $1.63\cdot10^{-4}$. This shows that our proposed approximation has a good accuracy. On the other hand, none of other state-of-the-art approximations is always tight.
Firstly, we can observe that, when $q$ and $p_0$ are large, \cite{2018Brown} significantly deviates from simulation results, such as $q=256,n=8,p_0=\{0.8,0.9\}$, $q=256,n=32,p_0=0.9$ and $q=256,n=128,p_0=0.96$. The maximum MSE of \cite{2018Brown} is occurred at $q=256,n=32,p_0=0.9$ and is equal to $1.30\cdot 10^{-3}$.
Secondly, we can observe that, when $n$ is small and $p_0$ is large, \cite{2019Chen} significantly deviates from simulation results, such as $q=\{2,256\},n=8,p_0=\{0.9,0.96\}$ and $q=\{2,256\},n=32,p_0=0.9$. The maximum MSE of \cite{2019Chen} is occurred at $q=2,n=8,p_0=0.96$ and is equal to $1.60\cdot10^{-2}$.
Finally, we can observe that \cite{2019Sehat} significantly deviates from simulation results except for a few cases. In particular, when $q$ and $p_0$ are large, \cite{2019Sehat} exhibits very low accuracy. The maximum MSE of \cite{2019Sehat} is occurred at $q=256,n=128,p_0=0.96$ and is equal to $7.75\cdot 10^{-1}$.

We now individually compare our proposed approximation with other state-of-the-art approximations. Firstly, it can be observed from Table \ref{table1} that, our proposed approximation is tighter than \cite{2018Brown} except for the cases of small $q$ and large $p_0$ and the case of $q=256,n=8,p_0=0.96$. Indeed, when $q$ is small and $p_0$ is large, the probability of the event that a set of rows sums to the zero vector is large. This leads to a high correlation between such events, which is significantly mitigated by \cite{2018Brown}.
As a result, \cite{2018Brown} is especially tight for small $q$ and large $p_0$. On the other hand, for small $q$ and large $p_0$, the inverse matrix is denser than the original matrix, which reduces the tightness of Lemma \ref{lemma3}. For above reasons, our proposed approximation is not as good as \cite{2018Brown} for small $q$ and large $p_0$.
Secondly, it can be observed that, with the only two exceptions of $q=\{2,256\},n=128,p_0=0.96$, our proposed approximation is tighter than \cite{2019Chen}. The reason is that the proposed approximation to $p(i,n)$ takes into account the correlation between the entries of a $n$-dimensional vector contained in a $i$-dimensional subspace. Furthermore, the proposed exact expression for the probability $P(n,m)$ as a function of $p(i,n)$ is of low complexity and needs no any approximations.
Finally, it can be observed that our proposed approximation is much tighter than \cite{2019Sehat} for all the cases considered. As mentioned in Section \ref{section1}, \cite{2019Sehat} does not take into account the correlation between linear dependencies of a matrix, which makes it particulary loose.

In order to comprehensively compare different approximations, for each case of Table \ref{table1}, we mark the best approximation with blue color and the worst approximation with red color.
We can observe that, our proposed approximation is best in $17$ of $24$ cases, \cite{2018Brown} is best in $6$ cases, \cite{2019Chen} is best in $9$ cases and \cite{2019Sehat} is best in $0$ cases, while our proposed approximation is worst in $0$ cases, \cite{2018Brown} is worst in $7$ cases, \cite{2019Chen} is worst in $1$ cases, and \cite{2019Sehat} is worst in $20$ cases.
In summary, in terms of the most best and the least worst, our proposed approximation outperforms \cite{2018Brown,2019Chen,2019Sehat}.

\section{Conclusion and future work}
\label{section5}
In this paper, we studied the perforamnce of multicast networks under SRLNC in terms of the decoding success probability which is given by the probability of a sparse random matrix over $F_q$ being full rank.
Based on the explicit structure of the RREF of a full row rank matrix and the product theorem, we derived a tight and closed-form approximation to the probability of a sparse random matrix over $F_q$ being full rank. This is in contrast with existing approximations which are recursive or not consistently tight. The accuracy of our proposed approximation was thoroughly assessed by Monte Carlo simulation for various configurations.
The simulation results show that our proposed approximation is of high accuracy regardless of the generation size, the number of coded packets, the field size and the sparsity, and tighter than the state-of-the-art approximations for a large range of parameters.

The proposed approximation can be used to derive other performance metrics, such as the rank distribution, the probability of partial decoding, the probability of all receivers recovering a generation, or the average number of coded packets required by a receiver to recover a generation or a portion of a generation.
We believe that, if the exact expression for each entry of the inverse matrix of a random matrix is known, the exact expression for the decoding success probability can be obtained by our proposed method. In the future, we will investigate the exact or tighter expression for each entry of the inverse matrix of a random matrix.

\appendix
\textit{Proof of Lemma \ref{lemma2}:} We prove the lemma by induction.
Because $F_q$ is a field, $g_1\neq 0$ and $X_1$ has a distribution (\ref{equation1}) with parameter $p_1$,
\begin{equation*}
Pr\{g_1X_1=t\}=Pr\{X_1=g_1^{-1}t\}=Pr\{X_1=t\}.
\end{equation*}
Therefore the lemma is true when $k=1$.
Assume that the lemma is true when $k=n$. Then
\begin{align*}
&Pr\{g_1X_1+\cdots+g_nX_n+g_{n+1}X_{n+1}=t~|~X_{n+1}=t_{n+1}\}\\
=&Pr\{g_1X_1+\cdots+g_nX_n=t+g_{n+1}X_{n+1}~|~X_{n+1}=t_{n+1}\}\\
=&Pr\{X_1+\cdots+X_n=t+g_{n+1}X_{n+1}~|~X_{n+1}=t_{n+1}\}\\
=&Pr\{X_1+\cdots+X_n+g_{n+1}X_{n+1}=t~|~X_{n+1}=t_{n+1}\}.
\end{align*}
According to total probability theorem,
\begin{align*}
&Pr\{g_1X_1+\cdots+g_nX_n+g_{n+1}X_{n+1}=t\}\\
=&Pr\{X_1+\cdots+X_n+g_{n+1}X_{n+1}=t\}.
\end{align*}
Similarly,
\begin{align*}
&Pr\{X_1+\cdots+X_n+g_{n+1}X_{n+1}=t~|~X_1=t_1,\cdots,X_n=t_n\}\\
=&Pr\{g_{n+1}X_{n+1}=t+X_1+\cdots+X_n~|~X_1=t_1,\cdots,X_n=t_n\}\\
=&Pr\{X_{n+1}=t+X_1+\cdots+X_n~|~X_1=t_1,\cdots,X_n=t_n\}\\
=&Pr\{X_1+\cdots+X_n+X_{n+1}=t~|~X_1=t_1,\cdots,X_n=t_n\}
\end{align*}
and
\begin{align*}
&Pr\{X_1+\cdots+X_n+X_{n+1}=t\}\\
=&Pr\{X_1+\cdots+X_n+g_{n+1}X_{n+1}=t\}\\
=&Pr\{g_1X_1+\cdots+g_nX_n+g_{n+1}X_{n+1}=t\}.
\end{align*}
Therefore, the lemma is also true when $k=n+1$. This completes the proof.

\textit{Proof of Lemma \ref{lemma3}:} It is easy to verify that the set $GL(i,q)$ of all nonsingular $i\times i$ matrices over $F_q$, with binary operation matrix multiplication, is a (nonabelian) group, called the general linear group \cite{2003Rotman}.
The number of $n\times m$ matrices over $F_q$ that have rank $r$ is (see \cite{1992Lint}, p338, Corollary of Th. 25.2)
\begin{equation*}
\begin{bmatrix}m\\r\end{bmatrix}_{q}
\sum_{k=0}^{r}(-1)^{r-k}\begin{bmatrix}r\\k\end{bmatrix}_{q}q^{nk+\binom{r-k}{2}}.
\end{equation*}
Therefore, the number of $i\times i$ matrices over $F_q$ that have rank $i$ is
\begin{equation*}
|GL(i,q)|=\sum_{k=0}^{i}(-1)^{i-k}\begin{bmatrix}i\\k\end{bmatrix}_{q}q^{ik+\binom{i-k}{2}}.
\end{equation*}
Since both $q$ and $i$ are finite, $GL(i,q)$ is a finite group. According to Lagrange theorem, the order of every element in $GL(i,q)$ is a factor of the order of $GL(i,q)$, therefore every element in $GL(i,q)$ has finite order.

For any $G\in GL(i,q)$, let $d(G)$ be the order of $G$. Then
\begin{equation*}
G^{d(G)}=I,
\end{equation*}
where $I$ is the identity matrix and $d(G)$ is the smallest positive integer such that $G^{d(G)}=I$,
and
\begin{equation*}
G^{-1}=G^{d(G)-1}.
\end{equation*}
Therefore, we have
\begin{equation*}
H^{d(H)}=I,
\end{equation*}
where $d(H)$ is the order of the random matrix $H$, and
\begin{equation*}
H^{-1}=H^{d(H)-1}.
\end{equation*}
Since it is not easy to calculate the power of a matrix and $d(H)$ is a random variable, reasonable approximation to the distribution of each entry of $H^{d(H)-1}$ is preferable. If $d(H)=1$, $H^{-1}=I=H$. If $d(H)\ge 2$, we approximate $d(H)$ by its lower bound, and therefore $H^{-1}=H^{d(H)-1}$ is approximated as $H$.
The lemma follows.

\ifCLASSOPTIONcaptionsoff
  \newpage
\fi

\bibliographystyle{IEEEtran}
\bibliography{myreferences}

\end{document}